\documentstyle[aps]{revtex}
\begin{document}
\title{Nagel Scaling and Relaxation in the Kinetic Ising Model on a n- Isotopic
Chain}
\author{L. L. Gon\c {c}alves\cite{lindberg}}
\address{Departamento de Fisica,\\
Universidade Federal do Cear\'{a},\\
Campus do Pici, C.P. 6030, 60451-970\\
Fortaleza, Cear\'{a}, Brazil}
\author{M. L\'{o}pez de Haro\cite{mariano}}
\address{Centro de Investigaci\'on en Energ\'{\i}a, UNAM, \\
Temixco, Morelos 62580, M\'{e}xico}
\author{J. Tag\"{u}e\~{n}a-Mart\'{\i}nez\cite{julia}}
\address{Centro de Investigaci\'on en Energ\'{\i}a, UNAM, \\
Temixco, Morelos 62580, M\'{e}xico}
\date{July 6, 2000}
\maketitle

\begin{abstract}
The kinetic Ising model on a n-isotopic chain is considered in the framework
of Glauber dynamics. The chain is composed of N segments with {\it n} sites,
each one occupied by a different isotope. Due to the isotopic mass
difference, the {\it n} spins in each segment have different relaxation
times in the absence of the interactions, and consequently the dynamics of
the system is governed by multiple relaxation mechanisms. The solution is
obtained in closed form for arbitrary {\it n}, by reducing the problem to a
set of $n$ coupled equations, and it is shown rigorously that the critical
exponent {\it z} is equal to 2. Explicit results are obtained numerically
for any temperature and it is also shown that the dynamic susceptibility
satisfies the new scaling (Nagel scaling) proposed for glass-forming
liquids. This is in agreement with our recent results (L. L. Gon\c{c}alves,
M. L\'{o}pez de Haro, J. Tag\"{u}e\~{n}a-Mart\'{\i }nez and R. B.
Stinchcombe, {\it Phys. Rev. Lett.} {\bf 84}, 1507 (2000)), which relate
this new scaling function to multiple relaxation processes.
\end{abstract}


The search for universal behavior in various physical model systems has been
one of the clues to uncover fundamental regularities of nature. About ten
years ago, a scaling hypothesis (the Nagel scaling) was proposed to describe
experimental work on dielectric relaxation in glass-forming liquids\cite
{Nagel}. This hypothesis was meant to replace the more usual normalized
Debye scaling, where the frequency is scaled with the inverse of the \
(single) relaxation time and the real and imaginary parts of the dynamic
susceptibility are then divided by their values at zero and one,
respectively. However, and in spite of its phenomenological success and
apparent ubiquity, its physical origin is as yet not quite well understood.

Very recently, we put forward what we believe to be the first {\it bona fide}
microscopic model in which the Nagel scaling is shown to arise\cite{nos2000}%
. In this previous work, we computed the magnetization, the dynamic critical
exponent $z$ and the frequency and wavevector dependent susceptibility of
the kinetic Ising model on an alternating isotopic chain with Glauber
dynamics\cite{Glauber}. Our results indicated that as soon as the two
relaxation times of this model became substantially different, the agreement
with the Nagel scaling improved significantly. The question remained of
whether the inclusion of more relaxation mechanisms would lead to the same
sort of results. It is this issue that we mainly want to examine in this
paper. Hence we present an extension of the alternating isotopic chain in
which rather than only two relaxation times, it is $n$ relaxation times
which are involved.

The model consists of a linear chain with $N$ segments, each one containing $%
n$ sites occupied by isotopes characterized by $n$ different spin relaxation
times. The Hamiltonian for the $l^{th}$ segment is the usual Ising
Hamiltonian given by

\begin{equation}
H_{l}=-J\sum\limits_{j=1}^{n}\sigma _{l,j}\sigma _{l,j+1},  \label{1}
\end{equation}

\noindent where $\sigma _{l,j}$ is a stochastic (time-dependent) spin
variable assuming the values $\pm 1$ and $J$ the coupling constant. Note
that if $n=2$, this model reduces to the alternating isotopic chain treated
in Ref. \cite{nos2000}. The configuration of the segment is specified by the
set of values $\left\{ \sigma _{l,1},\sigma _{l,2},...\sigma _{l,n}\right\}
\equiv \left\{ \sigma ^{l,n}\right\} $ at time $t$. This configuration
evolves in time due to interactions with a heat bath. We assume for the
segments the usual Glauber dynamics so that the transition probabilities are
given by

\begin{equation}
w_{li}(\sigma _{l,i})=\frac{1}{2}\alpha _{i}\left( 1-\frac{\gamma }{2}%
(\sigma _{l,i-1}\sigma _{l,i}+\sigma _{l,i}\sigma _{l,i+1})\right) ,
\label{2}
\end{equation}

\noindent where $\gamma =\tanh \left( 2J/k_{B}T\right) $, $k_{B}$ being the
Boltzmann constant and $T$ the absolute temperature, and $\alpha _{i}$ is
the inverse of the relaxation time $\tau _{i}$ of spin $i$ in the absence of
spin interactions.

The time dependent probability $P\left( \left\{ \sigma ^{l,n}\right\}
,t\right) $ for a given spin configuration satisfies the master equation

\begin{eqnarray}
\frac{dP\left( \left\{ \sigma ^{l,n}\right\} ,t\right) }{dt}
&=&-\sum\limits_{i=1}^{n}w_{li}\left( \sigma _{l,i}\right) P\left( \left\{
\sigma ^{l,n}\right\} ,t\right)  \nonumber \\
&&+\sum\limits_{i=1}^{n}w_{li}\left( -\sigma _{l,i}\right) P\left(
T_{li}\left\{ \sigma ^{l,n}\right\} ,t\right) ,  \label{3}
\end{eqnarray}
where $T_{li}\left\{ \sigma ^{l,n}\right\} \equiv \{\sigma _{l,1},\sigma
_{l,2},...\sigma _{l,i-1},-\sigma _{l,i},\sigma _{l,i+1},...\sigma _{l,n}\}$%
. The dynamical properties we are interested in, namely the magnetization,
the dynamic critical exponent and the susceptibility, require the knowledge
of some moments of the probability $P\left( \left\{ \sigma ^{l,n}\right\}
,t\right) $. Hence, we introduce the following expectation values defined as:

\begin{equation}
q_{li}\left( t\right) =\left\langle \sigma _{li}\left( t\right)
\right\rangle =\sum\limits_{\left\{ \sigma ^{l,n}\right\} }\sigma
_{l,i}P\left( \left\{ \sigma ^{l,n}\right\} ,t\right) ,  \label{4}
\end{equation}

\noindent where the sum runs over all possible configurations. Using the
master equation (3) and this definition of $q_{li}\left( t\right) $, one can
easily derive the result

\begin{eqnarray}
\frac{dq_{l,1}}{dt} &=&-\alpha _{1}\left( q_{l,1}-\frac{\gamma }{2}%
(q_{l-1,n}+q_{l,2})\right) ,  \nonumber \\
\frac{dq_{l,j}}{dt} &=&-\alpha _{j}\left( q_{l,j}-\frac{\gamma }{2}%
(q_{l,j-1}+q_{l,j+1})\right) ,(2\leq j\leq n-1)  \nonumber \\
\frac{dq_{l,n}}{dt} &=&-\alpha _{n}\left( q_{l,n}-\frac{\gamma }{2}%
(q_{l,n-1}+q_{l+1,1})\right) .  \label{5}
\end{eqnarray}

\noindent Introducing now the Fourier transform $\widetilde{q}_{Q,l}$ $%
(l=1,2,...n)$ defined by

\begin{equation}
\widetilde{q}_{Q,l}=\frac{1}{\sqrt{N}}\sum\limits_{l=1}^{N}\exp (iQld)q_{l,j}
\label{6}
\end{equation}

\noindent where $Q=\frac{2\pi m}{Nd}$ $(m=0,1,2...N)$, $d=na$, $a$ being a
lattice parameter and the vector $\Psi _{Q}$ given by

\begin{equation}
\Psi _{Q}=\left( 
\begin{array}{c}
\widetilde{q}_{Q,1} \\ 
. \\ 
. \\ 
. \\ 
\widetilde{q}_{Q,n}
\end{array}
\right) ,  \label{7}
\end{equation}

\noindent one can derive the following result

\begin{equation}
\frac{d\Psi _{Q}}{dt}={\bf M}_{Q}\Psi _{Q}  \label{8}
\end{equation}

\noindent where the matrix ${\bf M}_{Q}$ has a rather suggestive structure
given by

\begin{equation}
{\bf M}_{Q}=\left( 
\begin{array}{ccccccc}
-\alpha _{1} & \frac{\alpha _{1}\gamma }{2} & 0 & . & . & 0 & \frac{\alpha
_{1}\gamma }{2}e^{iQd} \\ 
\frac{\alpha _{2}\gamma }{2} & -\alpha _{2} & \frac{\alpha _{2}\gamma }{2} & 
0 & . & . & 0 \\ 
0 & \frac{\alpha _{3}\gamma }{2} & -\alpha _{3} & \frac{\alpha _{3}\gamma }{2%
} & 0 & . & 0 \\ 
. & . & . & . & . & . & . \\ 
. & . & . & . & . & . & . \\ 
. & . & . & 0 & \frac{\alpha _{n-1}\gamma }{2} & -\alpha _{n-1} & \frac{%
\alpha _{n-1}\gamma }{2} \\ 
\frac{\alpha _{n}\gamma }{2}e^{-iQd} & 0 & . & . & . & \frac{\alpha
_{n}\gamma }{2} & -\alpha _{n}
\end{array}
\right) .  \label{9}
\end{equation}

The solution to Eq. (8), which yields the magnetization, is straightforward,
namely

\begin{equation}
\Psi _{Q}\left( t\right) =e^{{\bf M}_{Q}t}\Psi _{Q}\left( 0\right) .
\label{10}
\end{equation}

The relaxation process of the wave-vector dependent magnetization is
determined by the eigenvalues of ${\bf M}_{Q}$ \ denoted as $\lambda _{lQ}$ $%
(l=1,2,...n)$ and which for $n\geq 5$ have to be computed numerically. The
same applies to the dynamic critical exponent $z$, which is obtained by
imposing the scaling relation $\tau _{Q}\sim \xi ^{z}f(\xi Q)$ for the
critical mode, $\lambda _{1Q},$ with the ($Q$-dependent ) relaxation time $%
\tau _{Q}=-\frac{1}{\lambda _{1Q}},$ in the region $T\rightarrow 0$ and $%
Q\rightarrow 0$ where $\lambda _{1Q}\rightarrow 0$, and the correlation
length $\xi $ is $\xi \sim \exp (2J/k_{B}T)$. This yields

\begin{equation}
\ln \left( -\frac{1}{\lambda _{1Q}}\right) =C+\frac{2Jz}{k_{B}T},  \label{11}
\end{equation}

\noindent where $C$ is an irrelevant constant. By plotting $\ln \left( -%
\frac{1}{\lambda _{10}}\right) $ vs. $\frac{2J}{k_{B}T}$ and considering the
limit $T\rightarrow 0$, we have checked analytically for $n=2$\cite{nos2000}
and numerically for $n=3,4$ and $5$ and various values of the $\alpha
^{\prime }s$ that $z=2$, so that this model belongs to the same universality
class of the uniform Ising chain.

Let us now introduce the spatial Fourier transform $\widehat{c}%
_{Q}(t^{\prime },t^{\prime }+t)$ of the time-dependent correlation defined
by 
\begin{equation}
\widehat{c}_{Q}(t^{\prime },t^{\prime }+t)=\frac{1}{nN}\sum\limits_{l=1}^{N}%
\sum\limits_{l^{\prime
}=1}^{N}\sum\limits_{i=1}^{n}\sum\limits_{j=1}^{n}e^{-iQdl}e^{iQdl\prime
}\left\langle \sigma _{l,i}\left( t^{\prime }\right) \sigma _{l^{\prime
},j}\left( t^{\prime }+t\right) \right\rangle .
\end{equation}
Then, the $t^{\prime }\rightarrow \infty $ limit of the temporal Fourier
transform of $\widehat{c}_{Q}\left( t^{\prime },t^{\prime }+t\right) ,$
denoted by $\widetilde{C}_{Q}(\omega ),$ is given by

\begin{equation}
\widetilde{C}_{Q}(\omega )=\lim_{t^{\prime }\rightarrow \infty }\frac{1}{%
2\pi n}\sum\limits_{i=1}^{n}\sum\limits_{j=1}^{n}\int_{-\infty }^{\infty
}\left\langle \widetilde{\sigma }_{-Q,i}\left( t^{\prime }\right) \widetilde{%
\sigma }_{Q,j}\left( t^{\prime }+t\right) \right\rangle \exp (-i\omega t)dt,
\label{12}
\end{equation}
with $\widetilde{\sigma }_{Q,m}=\frac{1}{\sqrt{N}}\sum\limits_{l=1}^{N}\exp
(iQld)\sigma _{l,m}$. After some lengthy and tedious but not very difficult
algebra, using eq. (\ref{4}) and the above definitions one may derive the
result

\begin{equation}
\widetilde{C}_{Q}(\omega )=\sum\limits_{l=1}^{n}\frac{g_{_{lQ}}}{i\omega
-\lambda _{lQ}}  \label{13}
\end{equation}
where

\begin{equation}
g_{lQ}=\sum\limits_{i=1}^{n}\sum\limits_{j=1}^{n}\sum\limits_{m=1}^{n}\frac{%
a_{ml}\overline{a_{lj}}}{n}\left\langle \widetilde{\sigma }_{-Q,i}\widetilde{%
\sigma }_{Q,j}\right\rangle _{eq}.  \label{14}
\end{equation}

Here, the static correlation functions $\left\langle \widetilde{\sigma }%
_{-Q,i}\widetilde{\sigma }_{Q,l}\right\rangle _{eq}$ are given by

\begin{equation}
\left\langle \widetilde{\sigma }_{-Q,i}\widetilde{\sigma }%
_{Q,l}\right\rangle _{eq}=\left\{ 
\begin{array}{cc}
\begin{array}{c}
\frac{1-u^{2n}}{1-2u^{n}\cos (nQ)+u^{2n}}
\end{array}
& 
\begin{array}{c}
(i=l)
\end{array}
\\ 
\begin{array}{c}
\\ 
u^{\left| l-i\right| }\frac{e^{iQn}u^{n}}{1-e^{iQn}u^{n}}+u^{-\left|
l-i\right| }\frac{e^{-iQn}u^{n}}{1-e^{-iQn}u^{n}} \\ 
+u^{\left| l-i\right| }
\end{array}
& (i\neq l)
\end{array}
\right.  \label{15}
\end{equation}
while $u=\tanh (\frac{J}{k_{B}T})$, $a_{ml}$ denotes the $(m,l)-$ element of
the matrix ${\bf A}$ formed with the eigenvectors of ${\bf M}_{Q}$ and $%
\overline{a_{lj}}$ the $(l,j)-$ element of the matrix ${\bf A}^{-1}$.
Finally, by using the fluctuation dissipation theorem\cite{Kubo}, the
response function $S_{Q}\left( \omega \right) $ turns out to be given by 
\begin{eqnarray}
S_{Q}\left( \omega \right) &=&\frac{1}{k_{B}T}\left( \frac{1}{n}%
\sum\limits_{i=1}^{n}\sum\limits_{j=1}^{n}\left\langle \widetilde{\sigma }%
_{-Q,i}\widetilde{\sigma }_{Q,l}\right\rangle _{eq}-i\omega \widetilde{C}%
_{Q}(\omega )\right)  \nonumber \\
&=&-\frac{1}{k_{B}T}\sum\limits_{l=1}^{n}\frac{\lambda _{lQ}g_{lQ}}{i\omega
-\lambda _{lQ}}.  \label{16}
\end{eqnarray}
\noindent It should be noted that the frequency dependent susceptibility is $%
\chi \left( \omega \right) \equiv k_{B}TS_{0}(\omega )(1-u)/(1+u)$. Thus,

\begin{equation}
\chi \left( \omega \right) =\frac{u-1}{u+1}\sum\limits_{l=1}^{n}\frac{%
\lambda _{l0}g_{l0}}{i\omega -\lambda _{l0}}  \label{17}
\end{equation}

If we set all the $\alpha ^{\prime }s$ equal in Eq. (17), {\it i.e. }we take
the uniform chain, then of course the resulting susceptibility has the
simple Debye form. In any case, its general structure is a linear
combination of $n$ Debye-like terms. In order to investigate whether the
Nagel scaling holds for this susceptibility, it is convenient to recall that
in the so-called Nagel plot the abscissa is $(1+W)\log _{10}\left( \omega
/\omega _{p}\right) /W^{2}$ and the ordinate is $\log _{10}\left( \chi
^{\prime \prime }(\omega )\omega _{p}/\omega \Delta \chi \right) /W$. Here, $%
\chi ^{\prime \prime }$ is the imaginary part of the susceptibility, $W$ is
the full width at half maximum of $\chi ^{\prime \prime }$, $\omega _{p}$ is
the frequency corresponding to the peak in $\chi ^{\prime \prime }$, and $%
\Delta \chi =\chi (0)-\chi _{\infty }$ is the static susceptibility. For the
sake of illustration in Figs. 1 and 2 we present Nagel plots for the cases $%
n=2$ (with $\alpha _{1}=1$ and $\alpha _{2}=2$), and $n=3$ (with $\alpha
_{1}=1$ and $\alpha _{2}=2$ and $\alpha _{3}=3)$, respectively, and
different values of the reduced (dimensionless) temperature $T^{*}\equiv
k_{B}T/2J$. By comparing the two figures, we can see that even for this
case, where the relaxation times are very close, there is a marked
improvement of the scaling in the low temperature limit with the increase of
the relaxation mechanisms. On the other hand, the Debye scaling, shown as
inset in the figures, presents an opposite behaviour. This result gives
further support to the hypothesis that the Nagel scaling is related to
multiple relaxation mechanisms, as discussed in our previous work [2].

\bigskip

\bigskip

One of us (L. L. Gon\c{c}alves) wants to thank the Centro de
Investigaci\'{o}n en Energia (UNAM) for the hospitality during his visit,
and the Brazilian agencies CNPq and FINEP for partial financial support. The
work has also received partial support by DGAPA-UNAM under projects IN104598
and IN117798.

\strut

{\bf Figure captions}

%
%\begin{figure}[htb]
%\begin{center}
%\epsfxsize=3.2in             %so many inches wide
%\leavevmode\epsfbox{Fig1.ps}
%\end{center}
%\caption{
%Nagel plot for $\alpha _{1}=1$ and $\alpha _{2}=2$ and for $%
%T^{\ast }=1,2,5,10$ and $100$. There is reasonable agreement with the
%scaling form for this choice except for low $T^{\ast }$.
%}%\label{}
%\end{figure}
%
%\begin{figure}[htb]
%\begin{center}
%\epsfxsize=3.2in             %so many inches wide
%\leavevmode\epsfbox{Fig2.ps}
%\end{center}
%\caption{
%The same as Fig. 1 but with the choice $\alpha _{1}=1$ , $\alpha
%_{2}=10$ and $T^{\ast }=5,10,50$ and $100$. The improvement in the agreement
%with the Nagel scaling is rather noticeable.
%}%\label{}
%\end{figure}
%
%\begin{figure}[htb]
%\begin{center}
%\epsfxsize=3.2in             %so many inches wide
%\leavevmode\epsfbox{Fig3.ps}
%\end{center}
%\caption{
%The same as Figs. 1 and 2 but for $\alpha _{1}=1$ , $\alpha
%_{2}=1000$ and $T^{\ast }=5,10,50$ and $100$. A plateau region is clearly
%present in this case.
%}%\label{}
%\end{figure}
%

Figure 1. Nagel plot for $n=2$ (with $\alpha _{1}=1$ and $\alpha _{2}=2$)
and for $T^{*}=5,7,10,50,100$ and $150$. There is reasonable agreement with
the scaling form for this choice except for low $T^{*}$. The insert contains
the plot of $\chi ^{\prime \prime }(\omega )/\chi ^{\prime \prime }(\omega
_{p})$ vs. $\omega /\omega _{p}$ in order to test the Debye-like behavior.

Figure 2. The same as Fig. 1 but with the choice $n=3$ ($\alpha _{1}=1$ , $%
\alpha _{2}=2$ and $\alpha _{3}=3$) and $T^{*}=5,7,10,50,100$ and $150$. The
improvement in the agreement with the Nagel scaling is rather noticeable,
while the opposite trend is observed with respect to the Debye scaling.

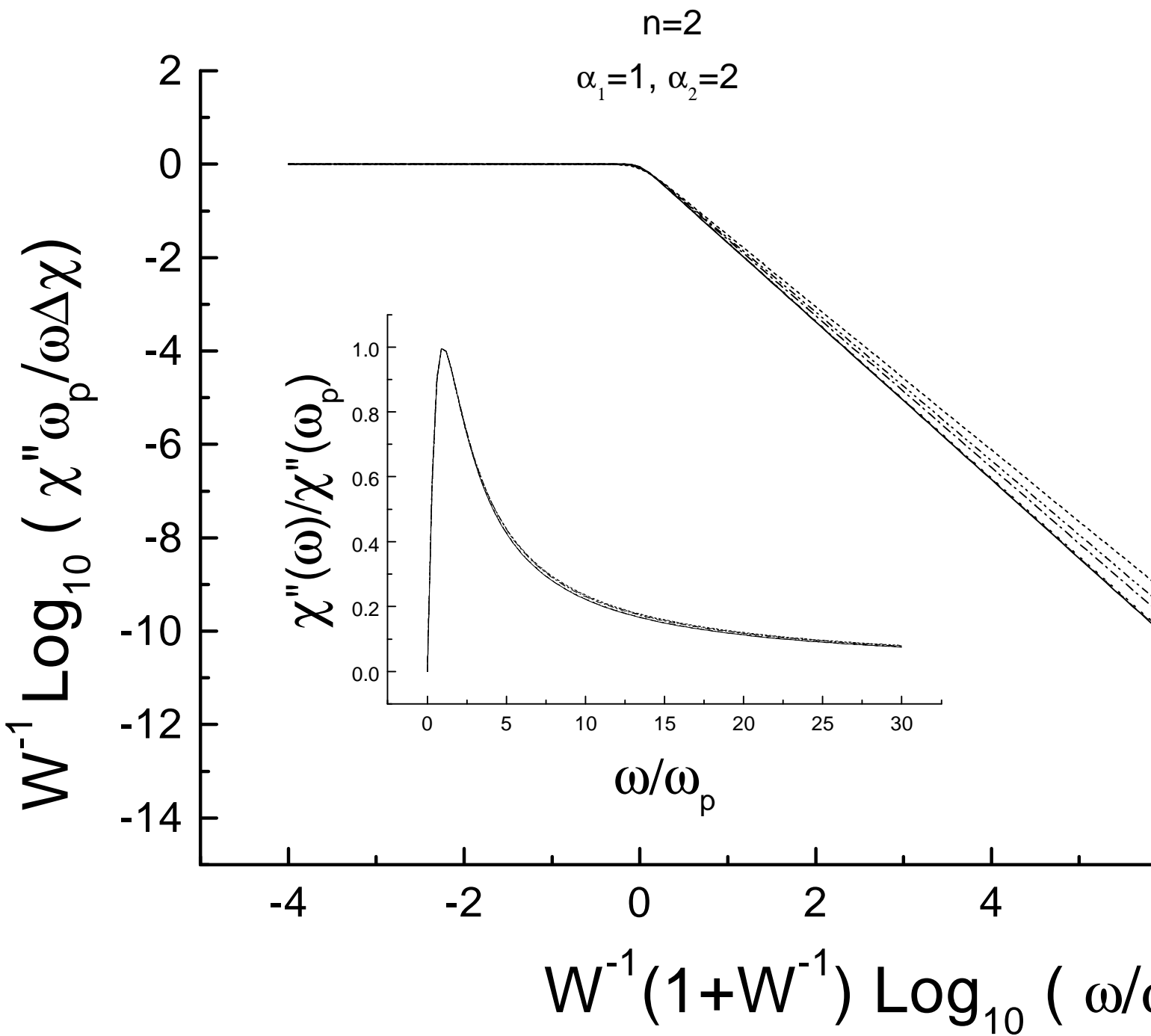
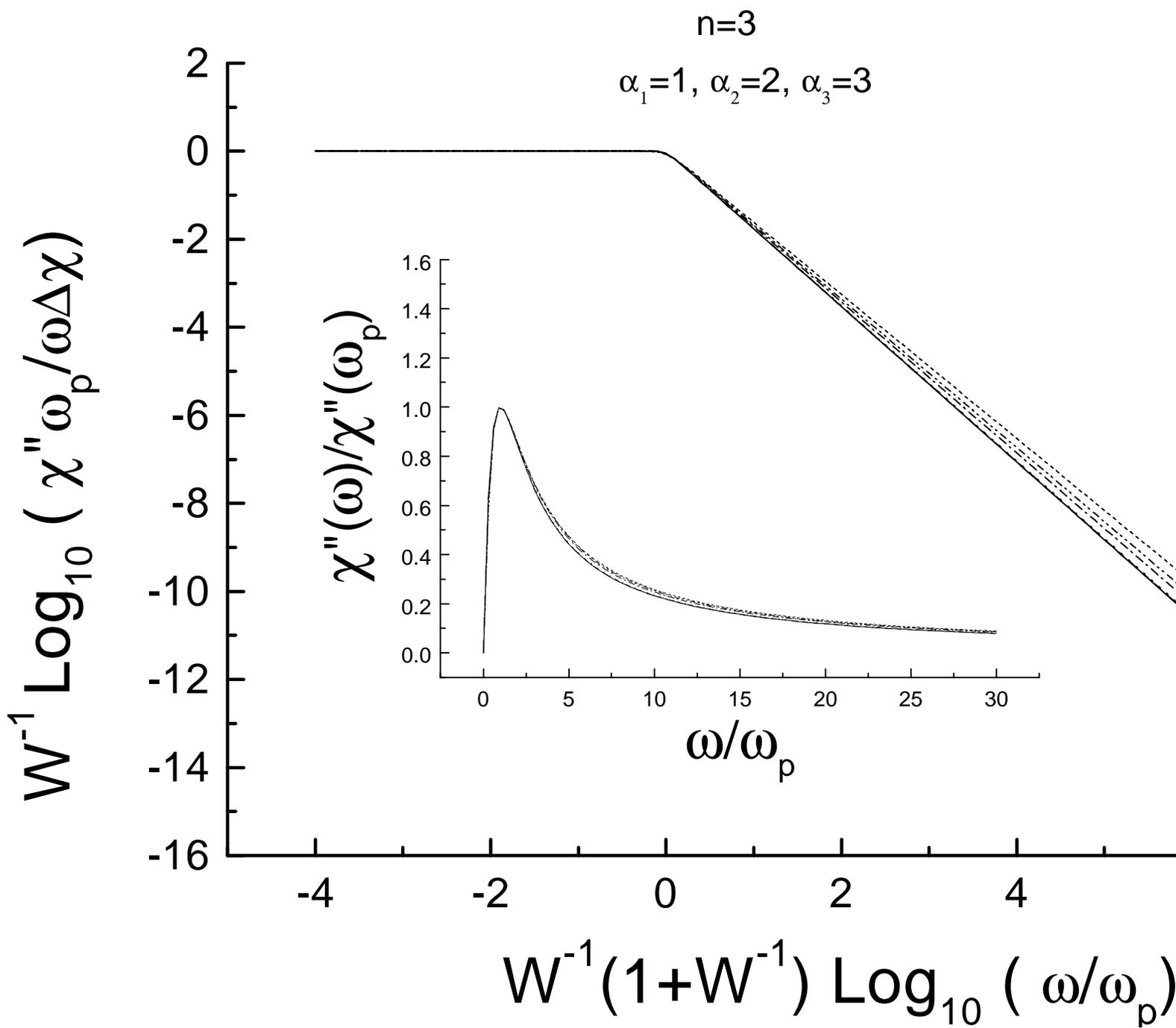
\end{document}